\documentclass[runningheads]{llncs}
\usepackage[T1]{fontenc}
\usepackage{graphicx}
\usepackage{amsmath,amssymb}
\usepackage{booktabs}
\usepackage{multirow}
\usepackage{xcolor}
\usepackage{hyperref}
\hypersetup{
    hidelinks,
    colorlinks=false,
    pdfborder={0 0 0}
}
\usepackage{orcidlink}
\newcommand{\tablefont}{\fontsize{7.5}{9.5}\selectfont}

\begin{document}

\title{An Imaging-Informed Reaction-Diffusion Model of Infarct Growth}
\author{\href{mailto:hussnainabbas14725@gmail.com}{Abbas, Muhammad Hussnain}\inst{1}\orcidlink{0009-0009-7775-2167} \and
        \href{mailto:michal.balcerak@uzh.ch}{Balcerak, Michal}\inst{2}\orcidlink{0009-0006-3137-7048} \and
        \href{mailto:asif.ahmad@giki.edu.pk}{Ahmad, Asif}\inst{3}\orcidlink{0000-0002-8299-1164} \and
        \href{mailto:bjoern.menze@uzh.ch}{Menze, Bjoern}\inst{2}\orcidlink{0000-0003-4136-5690} \and
        \href{mailto:ezequiel.delarosa@uzh.ch}{de la Rosa, Ezequiel}\inst{2}\orcidlink{0000-0002-9042-1962}}

\renewcommand{\thefootnote}{}
\footnotetext{Menze, Bjoern and de la Rosa, Ezequiel— Equal contribution.}

\authorrunning{Abbas et al.}
\institute{Faculty of Computer Science and Engineering, Ghulam Ishaq Khan Institute of Engineering Sciences and Technology, Topi, Pakistan
           \and
            Department of Quantitative Biomedicine, University of Zurich, Zurich, Switzerland
           \and
           Faculty of Basic Sciences, Ghulam Ishaq Khan Institute of Engineering Sciences and Technology, Topi, Pakistan}

\maketitle

\begin{abstract}
Predicting final ischemic infarct volumes from acute imaging is a cornerstone of personalized stroke management, yet current strategies remain polarized between uninterpretable machine learning architectures and overly detailed electrophysiological models that are intractable in clinical imaging settings. We bridge this clinical gap by introducing the first imaging-driven framework that parameterizes a Fisher-KPP reaction-diffusion partial differential equation (PDE) directly from clinical MRI. The continuous state variable $u(\mathbf{x},t)\in[0,1]$ models tissue damage, capturing the forward expansion of ionic stress through the extracellular space alongside a localized metabolic commitment to cell death gated within the baseline perfusion deficit ($T_{\max}>6$\,s). We evaluate this paradigm as an oracle: model and threshold parameters are fit to the 90-day outcome, so the results describe an upper bound on the model's feasibility and potential. On a subset of the ISLES 2017 dataset ($N=29$), \textcolor{black}{incorporating} biophysical propagation constraints yields a mean Dice score of $0.46 \pm 0.24$, compared to $0.25 \pm 0.21$ for standard rCBF thresholding. Ablations show that spatially varying the diffusion field with clinical perfusion maps improves on a reaction-only baseline and captures penumbral expansion, while keeping the model fully interpretable. This proof-of-concept shows that first-principles physics can capture part of ischemic lesion evolution directly on clinical scan grids, a step toward patient-specific biophysical infarct forecasting.

\end{abstract}

\section{Introduction}
Accurate prediction of the final ischemic stroke infarct volume is vital for tailoring personalized therapeutic strategies~\cite{albers2018thrombectomy}. In acute clinical settings, final infarct estimation directly guides decision-making by quantifying the potential benefit of mechanical thrombectomy, evaluating the efficacy of adjunctive neuroprotective agents, and anticipating malignant edema that may mandate decompressive hemicraniectomy. Furthermore, the final infarct volume serves as a robust surrogate biomarker for long-term functional outcomes~\cite{delarosa2025deepisles} and informs hospital resource optimization.

Despite its critical clinical utility, final infarct prediction remains an open challenge, with state-of-the-art machine learning models offering only modest predictive capabilities~\cite{delarosa2024isles}. While data-driven deep learning architectures excel at static ischemic lesion segmentation~\cite{delarosa2025deepisles,kamnitsas2017efficient}, they face fundamental limitations when modeling longitudinal tissue fate: they fail to capture the underlying biophysical mechanisms of infarct growth, lack interpretability, and require extensive annotated datasets that are difficult to curate for clinical outcomes. These constraints motivate the exploration of mechanistic, first-principles models capable of capturing complex infarct dynamics from small datasets while providing interpretable, physically grounded predictions.

Reaction-diffusion PDEs provide a well-established mathematical framework for 
modeling spatio-temporal biological processes. The Fisher--Kolmogorov--Petrovsky--%
Piskunov (Fisher-KPP) model~\cite{fisher1937wave,kolmogorov1937study,murray2002mathematical} captures progressive wavefront propagation and has been 
applied to tumor growth estimation~\cite{swanson2000quantitative,konukoglu2010image,balcerak2025gliodil} and neurodegeneration via amyloid-$\beta$ 
spreading in Alzheimer's disease~\cite{corti2024uncertainty}. Within the stroke 
domain, several groups have proposed mechanistic PDE models for ischemic damage 
propagation~\cite{chapuisat2008global,dronne2006modelling,dumont2013simulation}. 
These models operate at the level of ionic exchange dynamics, tracking 
transmembrane fluxes of Na$^{+}$, K$^{+}$, Ca$^{2+}$, and glutamate across 
intracellular, extracellular, and vascular compartments, and encoding processes 
such as spreading depolarizations and Na$^{+}$/K$^{+}$-ATPase pump failure. 
While biologically detailed, they require tens of parameters --- membrane 
conductances, ion pump rates, gap junction coupling constants --- that are not 
observable from clinical imaging and must be sourced from electrophysiology 
literature or in-vitro measurements. As a result, these models have not been 
driven by or evaluated against patient-specific clinical imaging 
data~\cite{rekik2012medical}. A mechanistic PDE formulation whose parameters are 
tractable and directly estimable from standard clinical imaging observations is 
absent from the literature.

In this work, we address this research gap with three key contributions. First, we introduce the first imaging-driven framework for modeling infarct growth using reaction-diffusion PDEs parameterized by and evaluated on clinical MRI data. Second, we perform a systematic validation of the proposed model using a sub-cohort of real acute stroke patients ($N=29$). Third, we conduct a comprehensive oracle experiment to establish the theoretical upper bounds and structural limitations of Fisher-KPP-based infarct prediction, demonstrating that mechanistic biophysical models offer a viable, interpretable alternative to domain-agnostic machine learning strategies.

\section{Methods}

\subsection{Proposed Framework --- Biological Interpretation}

We model the spatiotemporal evolution of ischemic tissue damage using a Fisher-KPP reaction-diffusion equation. The state variable $u(\mathbf{x},t) \in [0,1]$ represents fractional tissue damage, where $u=0$ denotes healthy tissue, $0<u<\theta$ denotes penumbral tissue at risk, and $u \ge \theta$ denotes irreversibly infarcted tissue. Critically, $u$ is not a statistical lesion probability; it is a physically grounded metabolic stress field that evolves according to biophysical principles. The core concept underlying this model is that infarction grows over time from the initial ischaemic core into the surrounding penumbra. The logistic reaction term drives local tissue commitment to infarction, while the diffusion term models the spatial propagation of ionic stress through the extracellular space.

We restrict the simulation domain $\Omega$ to the hypoperfused tissue territory ($T_{\max} > 6$ s), ensuring that the model operates only in regions at genuine perfusion risk, excluding tissue unimpacted by the arterial occlusion. The model covers the time window after stroke onset but does not simulate the occlusion event itself; it is confined to the already impacted area post-stroke. We refer to this as the \textit{perfusion-gated} Fisher-KPP model, as the simulation domain is strictly gated by the baseline perfusion deficit map.
The governing equation is:

\begin{equation}
\frac{\partial u}{\partial t} = \underbrace{\nabla\cdot\bigl(D(\mathbf{x})\nabla u\bigr)}_{\text{diffusion}} \,+\, \underbrace{\rho(\mathbf{x})\,u(1-u)}_{\text{reaction}},
\label{eq:fkpp}
\end{equation}
where $D(\mathbf{x})$ is the diffusion coefficient and $\rho(\mathbf{x})$ is the reaction rate.

The \textbf{diffusion term} models the lateral spread of ionic stress --- extracellular potassium, glutamate, and calcium imbalance --- through the brain's extracellular space. In ischemic tissue, cellular swelling reduces extracellular volume fraction from $\alpha\approx 0.20$ to $\alpha\approx 0.04$ and increases extracellular space tortuosity from $\lambda\approx 1.6$ to $\lambda\approx 2.2$, reducing effective diffusivity by approximately $1.9\times$~\cite{nicholson1998extracellular,sykova2008diffusion}. The \textbf{reaction term} models local metabolic commitment to infarction via logistic growth: once ATP depletion passes a critical threshold, the cell death cascade becomes irreversible. The logistic reaction term is monotone by design --- once $u\geq \theta$, tissue is committed to infarction and cannot recover within the model. Patients in this cohort with near-zero final infarction within the T$_{\max}>6$\,s territory fall outside the scope of this study (see Sect.~2.4 for the study design).

\subsection{Numerical Implementation}

The PDEs are solved via an explicit forward Euler scheme with a divergence-form spatial discretization using face-averaged coefficients~\cite{shashkov1996solving}. Numerical stability is maintained by computing a patient-specific time step via the Courant--Friedrichs--Lewy (CFL) condition~\cite{courant1928uber,mang2012biophysical}: $\Delta t = 0.45 / [2 D_{\max} \sum \Delta x_i^{-2}]$, where $D_{\max}$ is the maximum domain diffusivity and $\Delta x_i \in \{\Delta x, \Delta y, \Delta z\}$ are the voxel dimensions. For typical isotropic resolution ($\Delta x_i = 1.0$\,mm) and $D_{\max} \approx 0.10$\,mm$^2$/day, $\Delta t \approx 0.75$\,days, requiring $120$ steps over the $90$-day simulation horizon.

\subsection{Initial and boundary conditions}

The initial condition $u_0(\mathbf{x})$ represents the irreversibly damaged tissue at admission, generated from baseline ADC maps within the $T_{\max} > 6$\,s hypoperfusion mask. Voxels with ADC values $< 620 \times 10^{-6}$\,mm$^2$/s are considered core candidates~\cite{purushotham2015apparent,amukotuwa2019cerebral}, and the largest connected component is selected as the seed. To capture the graded, non-binary distribution of ischemia from the core to the periphery~\cite{fiehler2001adc}, $u_0(\mathbf{x})$ is continuously parameterized as $u_0(\mathbf{x}) = u_{\max} - (u_{\max} - u_{\min}) \times [(\text{ADC}(\mathbf{x}) - \text{ADC}_{\min}) / (\text{ADC}_{\max} - \text{ADC}_{\min})]$, clamped to $[u_{\min}, u_{\max}]$. We set $u_{\max} = 1.0$, $u_{\min} = 0.05$, $\text{ADC}_{\min} = 400 \times 10^{-6}$\,mm$^2$/s, and $\text{ADC}_{\max} = 620 \times 10^{-6}$\,mm$^2$/s. Crucially, avoiding a uniform $u_0 = 1.0$ prevents the immediate saturation of the logistic reaction term $\rho u(1-u)$, preserving early reaction dynamics within the seed interior. Finally, to confine the simulation to the anatomical parenchyma and prevent mass loss, we enforce a homogeneous Neumann boundary condition $\partial u / \partial \mathbf{n} = 0$ at the brain tissue interfaces.

\subsection{Experimental Design}
This study operates as an oracle simulation framework to characterize the capacity of reaction-diffusion equations to model ischemic lesion evolution, leveraging both admission multimodal MRI and 90-day follow-up ground truth for parameter fitting and evaluation. Rather than prospective prediction, our objective is to define the capabilities and structural limitations of these biophysical models via a systematic configuration search. 

We evaluate three PDE configurations, testing whether diffusivity $D$ and reaction rate $\rho$ are best modeled as homogeneous scalars or spatially varying fields:
$i)$ \textbf{Population-level scalars:} Isotropic values ($D = 0.1$~mm$^2$/day, $\rho = 0.1$~day$^{-1}$) optimized via leave-one-out cross-validation (LOOCV) to assess uniform growth dynamics. 
$ii)$ \textbf{Spatially varying diffusion:} A $T_{\max}$-modulated diffusivity field, $D(\mathbf{x}) = 0.01 + 0.09 e^{-\alpha T_{\max}(\mathbf{x})}$ with constant $\rho = 0.1$~day$^{-1}$ and $\alpha = 0.25$ (selected via a sensitivity sweep across $\alpha \in [0.1, 2.0]$). This formulation reflects the physiological mechanism where severe hypoperfusion induces cytotoxic edema, increasing cellular tortuosity and suppressing extracellular diffusion.
$iii)$ \textbf{Spatially varying reaction:} A $T_{\max}$-modulated metabolic vulnerability field, $\rho(\mathbf{x}) = 0.1 (1 - e^{-0.25 T_{\max}(\mathbf{x})})$ with constant $D = 0.1$~mm$^2$/day, testing the hypothesis that prolonged perfusion delays accelerate the local tissue commitment rate to infarction.

\textbf{Baselines.} The PDE configurations are compared against standard clinical deconvolution-based perfusion maps. Final infarct volumes are derived by thresholding $T_{\max}$ and rCBF using both standard clinical criteria ($T_{\max} > 6$\,s; $\text{rCBF} < 0.30$)~\cite{albers2018thrombectomy} and data-driven thresholds optimized via LOOCV. The same LOOCV procedure is used to select the PDE models' binarization threshold $\theta$.

\subsection{Dataset and Evaluation}
Simulations were evaluated on the ISLES 2017 dataset~\cite{winzeck2018isles}, which includes multimodal baseline MRI and 90-day follow-up infarct masks. Of the $N=44$ cases, 4 were excluded due to missing clinical data. To satisfy the model's assumption of continuous lesion progression from an observable seed, we excluded cases presenting embolic or extra-territorial strokes, or no baseline core within the perfusion deficit. This resulted in a cohort of 29 patients exhibiting at least minimal seed-to-target overlap ($\text{Dice} > 0.1$), thus retaining patients where a single and continuous diffusion-reaction growth process could describe infarct growth. The 15 excluded cases are mostly embolic, multi-territorial, or minimal-core presentations which may not be described through our model. The model was parameterized using co-registered baseline $T_{\max}$, ADC, and rCBF maps.

Performance was evaluated strictly within the $T_{\max} > 6$\,s hypoperfusion mask~\cite{olivot2009optimal} to restrict analysis to tissue at ischemic risk and exclude unrelated iatrogenic or embolic lesions. We report the Area Under the ROC Curve (AUC-ROC), Dice similarity coefficient, precision, and recall. The latter three metrics were computed after binarizing the continuous damage field $u(\mathbf{x}, T)$ at a threshold $\theta$ selected via the same LOOCV procedure described in Sect.~2.4, applied identically across models and baselines.

\begin{figure}[t]
  \centering
  \includegraphics[width=\textwidth]{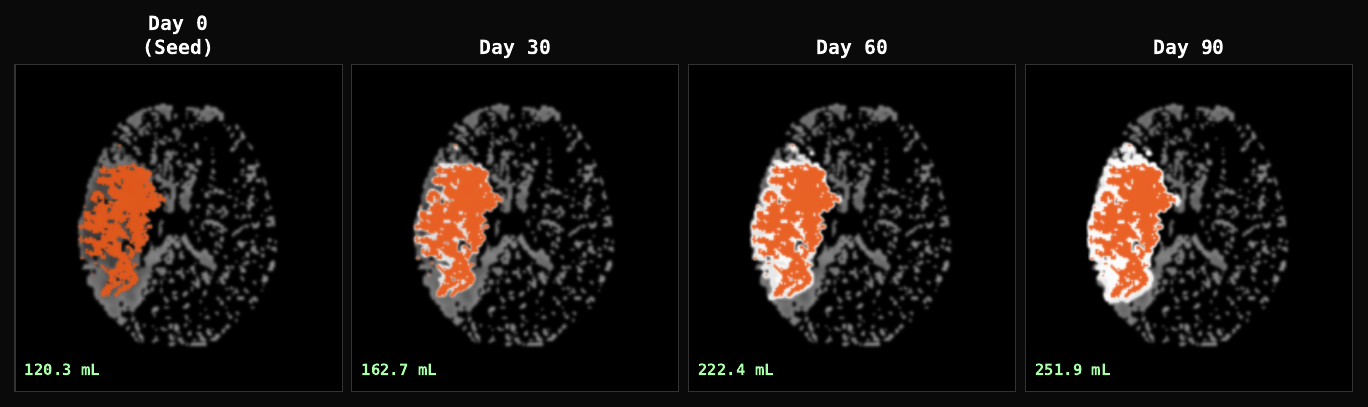}
  \caption{
    Simulated evolution of the proposed $D(x),\rho$ model for an ISLES'17 case, shown at four pseudo-time points of the internal simulation (Day 0, 30, 60, 90). The orange (white) mask represents the ADC-derived seed (simulated infarct growth).
  }
  \label{fig:progression_training4}
\end{figure}

\section{Results and Discussion}

\subsection{Model Performance}
Table~\ref{tab:results} summarizes baseline performance. The T$_{\max}$ clinical baseline (AUC-ROC $0.45 \pm 0.14$) shows a stark imbalance at the conventional 6\,s threshold, with near-perfect recall ($1.00 \pm 0.00$) but poor precision ($0.14 \pm 0.15$), reflecting systematic over-prediction from classifying the entire hypoperfusion territory as infarcted; the optimized threshold ($>9$\,s) does not meaningfully improve this. The rCBF baseline performs moderately better (AUC-ROC $0.70 \pm 0.10$, Dice $0.25 \pm 0.21$) with more balanced precision and recall, but static perfusion maps alone still lack the specificity to discriminate infarct from salvageable penumbra.

The reaction-only configuration achieved an AUC-ROC of $0.81 \pm 0.14$ and a Dice coefficient of $0.44 \pm 0.23$, showing that localized logistic growth from the graded ADC seed alone, with no spatial diffusion term, already reproduces most of the final lesion's shape. Introducing spatially homogeneous diffusion (Fixed $D$) raised the AUC-ROC to $0.89 \pm 0.11$, at the cost of lowering the Dice score to $0.41 \pm 0.23$; we analyze such a trade-off in the discussion section. Modulating the metabolic rate spatially while keeping diffusion uniform (Spatial $\rho$) produced metrics comparable to Fixed $D$ ($0.89 \pm 0.11$ AUC-ROC, $0.42 \pm 0.23$ Dice), so this variant offers no clear advantage over spatially uniform diffusion.

The Spatial $D(\mathbf{x})$ configuration, with a scalar reaction rate $\rho$, reached an AUC-ROC of $0.90 \pm 0.10$ and a Dice score of $0.46 \pm 0.24$, and the Joint Spatial model, which varies both fields, reached the highest overall performance ($0.91 \pm 0.10$ AUC-ROC, $0.47 \pm 0.23$ Dice). In Dice terms, given the small data cohort ($N=29$) and the large standard deviations obtained by most models, the marginal mean Dice improvement between reaction-only and the strongest spatial configurations might be interpreted as small given the spread of Dice scores. We therefore interpret the results of the Spatial $D(\mathbf{x})$, Spatial $\rho$, and Joint Spatial models with caution.
Comparing Spatial $D(\mathbf{x})$ (AUC-ROC $0.90 \pm 0.10$, Dice $0.46 \pm 0.24$) against the Joint Spatial model (AUC-ROC $0.91 \pm 0.10$, Dice $0.47 \pm 0.23$), the additional spatially-varying parameters in the Joint model yield only a marginal gain, so the added complexity may not be justified over Spatial $D(\mathbf{x})$ alone.
Analyzing results by seed fidelity clarifies the impact of the PDE diffusion operator. In patients whose acute ADC seed underestimated final infarct size ($N=11$), the Spatial $D(\mathbf{x})$ formulation reached a $0.12$ higher Dice score than the reaction-only variant. This suggests that spatially modulated diffusion is critical for capturing early ischemic progression and metabolic stress propagation through tissue at risk before overt structural damage manifests as acute ADC restriction.

\begin{table}[t]
\centering
\caption{Model performance. Values are mean $\pm$ standard deviation.}
\label{tab:results}
\tablefont
\begin{tabular}{lcccc}
\toprule
\textbf{Model} & \textbf{AUC-ROC} & \textbf{Dice} & \textbf{Precision} & \textbf{Recall} \\
\midrule
$T_{\max}$ (Fixed, $>6$\,s)              & $0.45\pm0.14$ & $0.22\pm0.22$ & $0.14\pm0.15$ & $\mathbf{1.00\pm0.00}$ \\
$T_{\max}$ (Optimized, $>9$\,s)          & $0.45\pm0.14$ & $0.23\pm0.22$ & $0.16\pm0.18$ & $0.79\pm0.16$ \\
rCBF (Fixed, $<0.30$)                    & $0.70\pm0.10$ & $0.25\pm0.21$ & $0.22\pm0.21$ & $0.44\pm0.25$ \\
rCBF (Optimized, $<0.33$)                & $0.70\pm0.10$ & $0.25\pm0.22$ & $0.22\pm0.21$ & $0.50\pm0.25$ \\
\hline
Reaction-only                            & $0.81\pm0.14$ & $0.44\pm0.23$ & $\mathbf{0.48\pm0.28}$ & $0.64\pm0.32$ \\
Fixed $D$                                & $0.89\pm0.11$ & $0.41\pm0.23$ & $0.38\pm0.28$ & $0.77\pm0.24$ \\
Spatial $\rho$                           & $0.89\pm0.11$ & $0.42\pm0.23$ & $0.39\pm0.28$ & $0.76\pm0.24$ \\
Spatial $D$                              & $0.90\pm0.10$ & $0.46\pm0.24$ & $0.43\pm0.27$ & $0.75\pm0.27$ \\
Joint Spatial                            & $\mathbf{0.91\pm0.10}$ & $\mathbf{0.47\pm0.23}$ & $0.46\pm0.27$ & $0.73\pm0.27$ \\
\bottomrule
\end{tabular}
\end{table}

\subsection{Ablations}
Table~\ref{tab:ablations} presents the results of the ablation experiments.
Reaction-only achieves $0.81\pm0.14$ AUC, demonstrating 
that logistic growth from the graded ADC seed already captures a substantial 
portion of the lesion without any spatial diffusion. Adding fixed homogeneous 
diffusion (Fixed $D,\rho$, $0.89\pm0.11$ AUC-ROC) reduces Dice to $0.41\pm0.23$, 
as isotropic diffusion spreads damage into the penumbra without constraint, 
inflating false positives. Spatially varying diffusion ($D(\mathbf{x})$, 
$0.90\pm0.10$ AUC-ROC) corrects this by suppressing wave propagation in severely 
hypoperfused regions where cytotoxic edema increases extracellular 
tortuosity~\cite{sykova2008diffusion}, improving over reaction-only by $0.09$ 
AUC. Spatially varying $\rho(\mathbf{x})$ with fixed $D$ ($0.89\pm0.11$) 
offers no meaningful gain, suggesting the metabolic vulnerability gradient is 
already captured by diffusion modulation alone. The Joint Spatial model adds 
only $0.01$ AUC ($0.91\pm0.10$) at the cost of increased complexity; we 
therefore select $D(\mathbf{x}),\rho$ as the proposed model. For seed 
initialisation, the single-point centre-of-mass seed performed worst 
($0.85\pm0.14$), while the ADC-based multi-point seed ($0.90\pm0.10$) 
outperforms the rCBF-based alternative ($0.86\pm0.08$), consistent with ADC 
being a more direct marker of irreversible cytotoxic injury. The decay 
constant $\alpha=0.25$ yields the best AUC ($0.90\pm0.10$), and AUC stays within $0.01$ of this value across $\alpha \in [0.10, 0.50]$, dropping only slightly beyond $\alpha=1.00$ ($0.89\pm0.12$), so the model is not sensitive to the exact choice of $\alpha$ within this range. It is similarly robust to the imaging 
source for $D(\mathbf{x})$: T$_{\max}$-, ADC-, and rCBF-based 
parametrizations land within $0.01$ AUC of each other.

\subsection{Discussion}
In this oracle-setting study, PDE-based reaction-diffusion models substantially outperformed static perfusion baselines across all metrics (Table~\ref{tab:results}). Given the minimal performance gain of the joint spatial formulation over the Spatial $D(\mathbf{x})$ model (Table~\ref{tab:ablations}), we select $D(\mathbf{x})$, $\rho$ as the primary architecture, since it strikes an optimal balance between predictive accuracy and model parsimony, avoiding unnecessary parametric complexity. Robustness to the imaging source used to parametrize $D(\mathbf{x})$ is also clinically relevant, as it implies that the model may not depend on a specific scan modality for estimating the diffusion operator and, hence,  broadening its potential to diverse imaging modalities and acquisition protocols.

Our analysis reveals three patient categories that systematically bound model performance. First, cases with minimal final infarct volumes are disproportionately sensitive to spatial misalignment between the initial ADC seed and the follow-up lesion. Second, patients presenting with no detectable baseline ADC core fall outside the model's structural operating assumptions entirely, as no anatomical seed can be generated. Third, patients who achieve successful recanalization (TICI 2b--3) expose a structural boundary of the model: the Fisher-KPP framework models continuous lesion expansion under sustained ischemia and cannot capture tissue salvage or arrest following timely reperfusion~\cite{fiehler2002adc}. Future iterations could resolve this by integrating time-to-treatment kinetics to dynamically truncate or modulate the simulation upon recanalization. Finally, the primary limitations of this proof-of-concept study remain the small cohort size ($N=29$) and the absence of direct comparisons against domain-agnostic machine learning models, highlighting the need for benchmark comparisons and validation on larger, multi-center datasets.

\begin{table}[t]
\centering
\caption{Ablation results (mean $\pm$ SD). $D(\mathbf{x})$ and $\rho(\mathbf{x})$ denote spatially varying fields. ADC (CoM) is a single-point, center-of-mass seed. }
\label{tab:ablations}
\tablefont
\setlength{\tabcolsep}{4.5pt}
\begin{tabular}{lcclcclcc}
\toprule
\multicolumn{2}{c}{\textbf{PDE Config.}} & \phantom{i} & \multicolumn{2}{c}{\textbf{Seed Init.}} & \phantom{i} & \multicolumn{2}{c}{\textbf{Alpha ($\alpha$)}} \\
\cmidrule{1-2} \cmidrule{4-5} \cmidrule{7-8}
Model & AUC && Type & AUC && $\alpha$ & AUC \\
\midrule
Reaction only ($\rho$, $D=0$) & $0.81 \pm 0.14$ && ADC (CoM) & $0.85 \pm 0.14$ && $0.10$ & $\mathbf{0.90} \pm \mathbf{0.10}$ \\
$\rho(\mathbf{x})$, $D$   & $0.89 \pm 0.11$ && rCBF      & $0.86 \pm 0.08$ && $0.25$ & $\mathbf{0.90} \pm \mathbf{0.10}$ \\
$D(\mathbf{x})$, $\rho$   & $0.90 \pm 0.10$ && ADC       & $\mathbf{0.90} \pm \mathbf{0.10}$ && $0.50$ & $0.90 \pm 0.11$ \\
$D(\mathbf{x})$, $\rho(\mathbf{x})$ & $\mathbf{0.91} \pm \mathbf{0.10}$ &&       &                 && $1.00$ & $0.89 \pm 0.12$ \\
           &                 &&       &                 && $2.00$ & $0.89 \pm 0.12$ \\
\bottomrule
\end{tabular}
\end{table}

\section{Conclusions}
In this work, we presented the first imaging-driven Fisher-KPP reaction-diffusion framework to model ischemic stroke lesion expansion directly on clinical MRI grids. By formulating an oracle simulation environment, we established the feasibility, predictive boundaries, and physiological constraints of continuous biophysical modeling in acute stroke. Our findings demonstrate that a parsimonious biophysical model initialized solely from standard acute MRI effectively captures spatiotemporal lesion dynamics, outperforming standard clinical perfusion baselines. This provides a strong case for treating first-principles partial differential equations as an interpretable, physically grounded complement to domain-agnostic machine learning architectures.

Crucially, our results reveal that model performance is driven by physiological regularization rather than parametric expansion. Spatially modulating the diffusivity field $D(\mathbf{x})$ via clinical perfusion maps significantly improves lesion characterization, whereas adding spatial heterogeneity to the reaction rate $\rho(\mathbf{x})$ provides marginal benefit beyond what diffusion alone captures. Importantly, the upper bounds established in this oracle evaluation are inherently constrained by the specific mathematical parameterizations chosen for these spatially varying fields. Future work should explore a wider array of functional parameterizations to determine whether alternative formulations can push these upper performance boundaries further and yield even more accurate biophysical models of infarct growth.

Finally, the model’s performance boundaries in successfully recanalized patients (TICI 2b--3) stem directly from its monotonic growth assumption under sustained ischemia. Rather than a fundamental deficiency, this defines a clear clinical scope that outlines the roadmap for translation. Moving from this oracle proof-of-concept to true prospective prediction requires tackling the core challenge of inferring patient-specific PDE parameters solely from acute admission imaging and baseline clinical data. Addressing parameter estimation from acute imaging alone, alongside treatment timing, is a key step toward evaluating whether biophysical frameworks can ultimately support personalized stroke forecasting.

\end{document}